\documentclass[twocolumn,showpacs,amsmath,amssymb]{revtex4}
\usepackage{epsfig}
\usepackage{dcolumn}
\usepackage{bm}
\begin{document}
\title{Raman thresholds and rigid to floppy transitions 
in calcium silicate glasses}
\author{M. Micoulaut$^1$, M. Malki$^{2,3}$, P. Simon$^2$ and A. Canizares$^2$}
\affiliation{$^1$Laboratoire de Physique Th{\'e}orique des Liquides,
Universit{\'e} Pierre et Marie Curie, Boite 121,
4, Place Jussieu, 75252 Paris Cedex 05, France\\
$^2$ Centre de Recherche sur les Mat\'eriaux \`a Haute Temp\'erature, 1D,
Avenue de la Recherche Scientifique\\45071 Orl\'eans Cedex 02 France\\
$^3$Ecole Polytechnique de l'Universit\'e d'Orl\'eans, 
8 rue L\'eonard de Vinci, 45072 Orl\'eans 
Cedex 02 France}
\date{\today}
\begin{abstract}
Alkaline earth silicate glasses $xCaO-(1-x)SiO_2$ exhibit a well marked
threshold in Raman lineshapes which can be related to the onset of network
rigidity as the concentration of calcium oxide $x$ is decreased. The present
results are analyzed by constraint counting algorithms and more deeply
characterized by a size increasing cluster approximation that
allows to perform Maxwell mechanical constraint counting beyond the usual 
mean-field treatment. This permits to discuss under which structural conditions
an elastic intermediate phase can be obtained.
\end{abstract}
\pacs{61.20N-81.20P}
\maketitle
\section{Introduction}
Magmatic liquids are the principal agents of mass and heat transfer in
the Earth and terrestrial planets and intensive research has been
accomplished to understand the processes of mass or energy transfer
with respect to melt or structural properties \cite{r1,r2}. 
Viscosity, thermal or
electrical conductivity are indeed directly related to the structure of
silicate melts which furthermore control the temperature behavior of magmas
and their formation or crystallization \cite{r3,r4}.
\par
In this context, calcium silicates of chemical formula $xCaO-(1-x)SiO_2$
have received little attention as
compared to the corresponding alkaline systems so that much of their
properties over the whole calcium glass formation range are still object of 
active research.
Most studies have indeed focused on the $x=0.50$ molar concentration which
corresponds to the crystalline wollastonite composition \cite{r5}.
Several studies have stressed the special role played by the calcium atom
which acts as a modifier in the silicate networks \cite{r6} and leads 
to a global
increase of the density from $2.75~g.cm^{-3}$ at $x=0.38$ to $2.97~g.cm^{-3}$
at the $x=0.60$ pyrosilicate composition \cite{Doremus,Doweidar}.
Extensive studies have been performed to understand the miscibility limits
in this system and in other alkaline earth silicates \cite{r7,r8}.
\par
A special attention has been devoted to the coordination number of the
calcium atom using EXAFS \cite{r9} and X-ray spectroscopy \cite{r10}.
More recently, Yannopoulos and co-workers have been studying inelastic light
scattering of calcium silicates around the $x=0.50$ composition by 
polarized Raman spectroscopy \cite{r11}. 
The results with composition show a marked
change in behavior around $47\%$ calcium. Specifically, intensity ratios of
particular Raman lines as well as the Boson peak frequency present a sharp
jump at this concentration. On the other hand, a line at $606~cm^{-1}$ shows
an abrupt increase but there remain some uncertainety about its 
attribution to the so-called $D_2$ ring line \cite{r12}. 
However, the 
conclusion obviously suggests the presence of a transition that has not been
characterized by the authors.
\par
In this work, we show from the investigation of Raman spectroscopy that a very
particular elastic state is reached in the glass when the concentration of
calcium $x$ equals $47\%$. This threshold is identified with
a rigid to floppy transition from Maxwell mechanical constraint counting 
and suggests that elastic transitions can
take place in calcium silicate glasses as in chalcogenides \cite{r13,r14}. 
In the latter, the addition of cross-linking units such as germanium or 
silicon into a 
basic network made of two-folded chalcogenide atoms 
(sulphur, selenium) constraints indeed internal degrees of freedom
of the network by increasing the number of bond-bending and 
bond-stretching forces that can lead to a very peculiar situation when 
the number of constraints per atom equals the number of degrees of freedom
\cite{r15}.
It has been identified by M.F. Thorpe from numerical simulations on amorphous
silicon as being a floppy to rigid transition which is characterized by the
vanishing of the number of normal (floppy) modes of the dynamical matrix
\cite{r16,r17}. The glass transition of such systems undergoing a rigid to 
floppy transition is also substantially affected \cite{Naumis0}.
Understanding to what extent the methods used with success for the 
chalcogenides can be applied to oxide glasses is therefore not only of 
fundamental interest but also an attractive perspective for applied purposes
\cite{r18}.
\par
We first display the results of the Raman analysis and 
the deconvolution of the spectra which show a marked change 
in behaviour for some modes at $x=0.47$. 
Next, we apply on this system mean-field
Maxwell mechanical constraint counting that permits to compute the
concentration $x_c$ at which the fraction of zero frequency (floppy)
modes vanishes \cite{r16}. 
We improve the approach in Section IV by analyzing the present system
with size increasing cluster approximations (SICA) that infer the effect
of medium-range order on the nature and the location of the transition which
is found to be in harmony with experimental findings. Finally, we show which 
conditions in terms of medium range order have to be satisfied in order 
to obtain an intermediate self-organized phase \cite{r19} that is bounded by a 
rigidity and a stress
transition \cite{r20}, in close correspondance with the chalcogen analog.
\section{Experimental results}
\subsection{Sample preparation}
The samples were prepared by mixing pre-dried $SiO_2$ (99.99 $\%$) and
$CaCO_3$ (99.95 $\%$) powders in the correct proportions. For each
sample, the mixture was melted in a platinum crucible at $1650^{o}$C for 
two hours, and quenched by placing the bottom of the crucible 
in cold water. The samples were then annealed at a temperature around $760^oC$
for five hours and cooled slowly to room temperature. The glasses were 
transparent and free of crystallization as confirmed by the absence of Bragg
peaks in XRD spectra. Chemical microanalysis using the energy-dispersive
X-ray (EDX) performed on optically polished glasses showed a very 
low departure (less than 0.4 mol of Ca) between the theoretical and 
the real composition for all the studied samples. 
The glass 
transition temperatures were determined with a differential scanning 
calorimeter Setaram DSC-1600 at a heating rate $10^{o}C/min$. These
values (see Table \ref{table}) are slightly larger than those reported 
by Shelby\cite{Shelby}
using the dilatometric technique, but they exhibit the same global trend
\begin{table}[t]
\caption{\label{table} Glass transition temperatures of the $(1-x)SiO_2-xCaO$ 
system.}
\begin{ruledtabular}
\begin{tabular}{cccccc}
Concentration $x$&42&44&47&50&53 \\
$T_g$ [$^oC$]&769&769&770&781&795 \\
\end{tabular}
\end{ruledtabular}
\end{table}
\subsection{Raman scattering}
Raman spectra were obtained on a Jobin-Yvon T64000 spectrometer, with CCD 
detection, and through a BX40 Olympus microscope (objective x100).  The 
excitation wavelength was the $514.532~nm$ argon line of a Coherent Innova 
70 Spectrum laser. The power was typically 300mW at the laser output 
($20 mW$ on the sample). The low-frequency part range ($<500cm^{-1}$) was 
obtained in triple substractive mode (gratings $1800gr/mm$). For the 
high frequency part (above $300~cm^{-1}$), the spectrometer was in simple 
monochromator configuration with Notch filter (grating $600gr/mm$). 
This allows a sufficient recovering spectral range for merging the 
spectra. All the data on the different compositions were collected 
in exactly the same conditions, one following each other, in order 
to be the most confident possible in any variation between subsequent spectra.
\begin{figure}
\begin{center}
\includegraphics[width=0.9\linewidth]{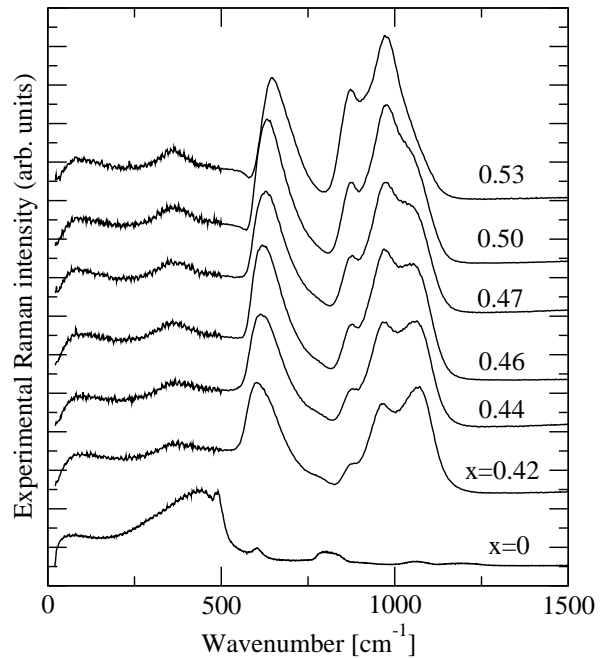}
\caption{\label{raman} Raman lineshapes for various calcium concentrations. The
low wavenumber range was acquired in triple substractive mode (see text) which
explains the higher noisy level. The spectrum of silica at the bottom of the
figure is displayed for comparison.}
\end{center}
\end{figure}
\subsection{Peak deconvolution}
The experimental spectra were first corrected from first-order 
Bose-Einstein factor, and also from scattering law  ($I\simeq\nu^4$). The latter 
(a second-order correction compared to Bose-Einstein factor) is not of 
primary importance but is theoretically needed to extract the Raman 
susceptibility, which is the relevant physical quantity which 
can be 
used to extract information on dynamics. The Bose Einstein reduction 
does not include the prefactor $1/\omega$  generally used for describing 
the low-frequency part of Raman spectra in glasses \cite{r24}, according to 
the Shuker-Gammon formalism \cite{r25}. Examination of the bare spectra 
shows that 
the main variations upon increasing $Ca$ content from $0.42$ to $0.53$ lie 
at wavenumbers above $300cm^{-1}$ and then justify to focus attention on 
this spectral range, dominated by intratetrahedron or $Si-O-Si$ motions. 
The boson peak reflects order at larger distances. After this, the 
spectra were reconstructed with the FOCUS software \cite{Simon} , by 
using a log-normal law for the boson peak at low frequencies, and 
gaussian shapes for higher frequency ones. The low-frequency (boson 
peak) part was fitted to insure a correct description of the mid-frequency 
range, due to its long tail, but is clearly out of the scope of the 
present paper. Moreover, recent hyperRaman measurements in silica tend to show that the Boson peak responsable for the excess density of states is directly
observed in hyperRaman and inelastic X-ray scatteing, meaning that the
low-frequency Raman component implies other degrees of freedom and is then 
much more complicated to interpret \cite{Vacher}.
For higher frequency modes, gaussian shapes gave a 
better reconstruction of the spectra than lorentzian  or damped 
harmonic oscillator ones 
\begin{figure}
\begin{center}
\includegraphics[width=0.9\linewidth]{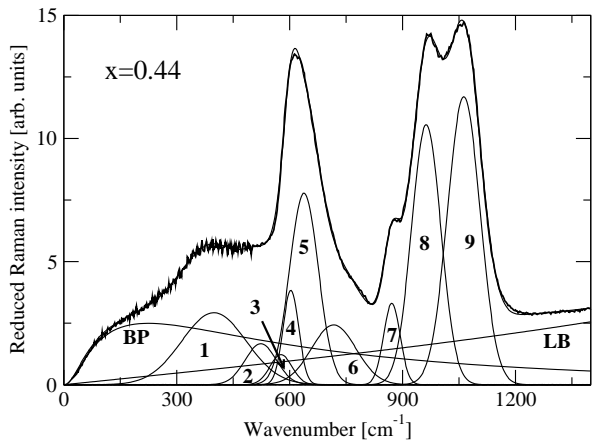}
\end{center}
\caption{\label{ramanfit} Fits of the Raman spectra of the $x=0.44$ sample. 
The numbers (1-9) correspond to the various gaussian deconvolutions of the 
spectra (see text for details). BP is the Boson peak and LB the luminscence
background.}  
\end{figure}
\section{Discussion}
\subsection{Spectra description}
\begin{figure}
\begin{center}
\includegraphics[width=0.9\linewidth]{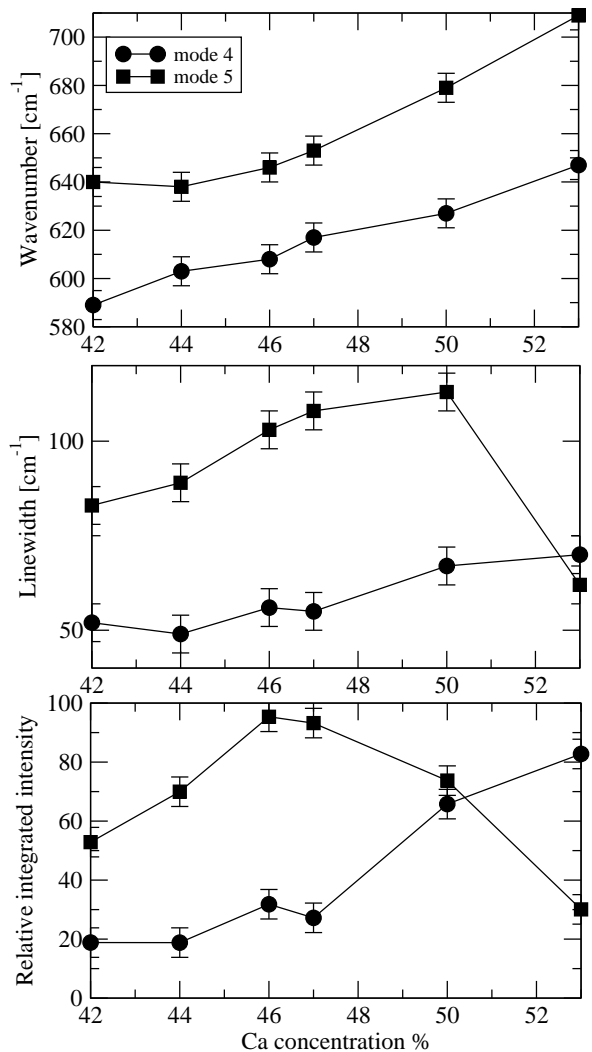}
\caption{\label{raman2}Compositional behaviour in wavenumber, 
linewidth and relative integrated intensity for the lines 4 and 5
of the Raman spectra in the $(1-x)SiO_2-xCaO$ system.} 
\end{center}
\end{figure}

\begin{figure}
\begin{center}
\includegraphics[width=0.9\linewidth]{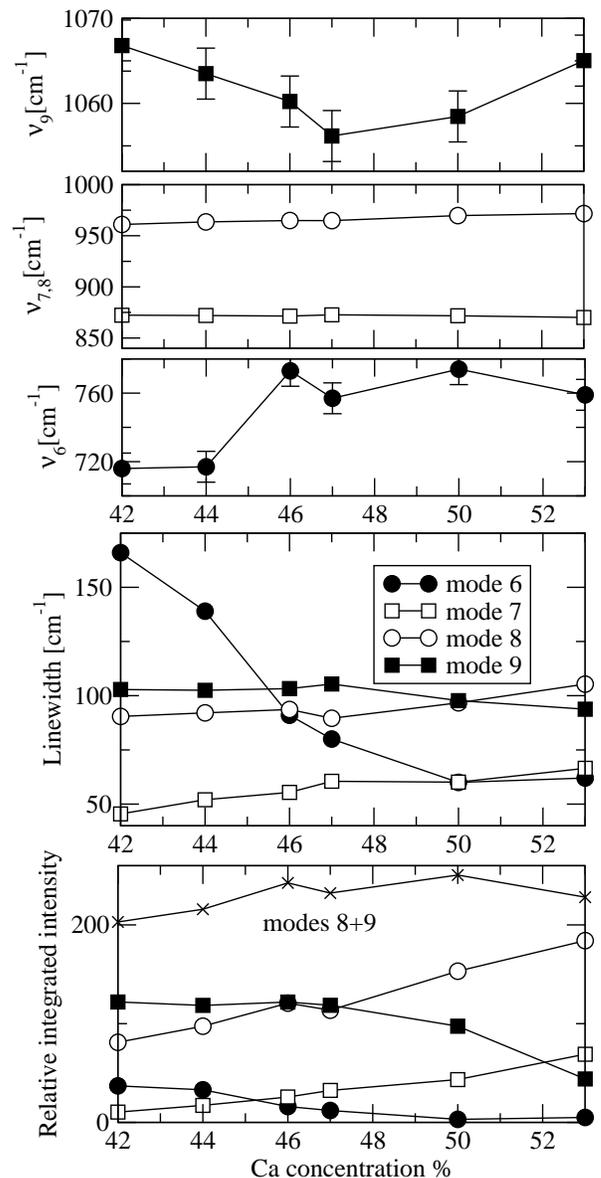}
\caption{\label{raman1} Compositional behaviour in wavenumber, 
linewidth and relative integrated intensity for different lines (6-9)
of the Raman spectra in the $(1-x)SiO_2-xCaO$ system. The error bars are of 
the size of the symbols except for mode 6 and 9 wavenumbers.}
\end{center}
\end{figure}

Figure \ref{raman} displays the experimental Raman data for the six $Ca$ 
concentrations $0.42$ to $0.53$. Figure \ref{ramanfit} exhibits the typical 
agreement between experimental and fitted spectra (Bose-Einstein and 
scattering corrected). Results of the fitting procedure are displayed 
in the different panels of Figure \ref{raman2} and \ref{raman1} 
(line wavenumbers, linewidths and integrated intensity). 
The whole concentration range was described with 9 modes 
(plus the boson peak) : 3 modes in the range below $550~cm^{-1}$, 3 modes 
to describe the sharp and assymetric peak near $600~cm^{-1}$, and 3 modes 
for the 
higher-frequency group around $1000~cm^{-1}$. A broad luminescence background 
was needed to describe the intensity increase up to high relative 
frequencies, a feature that can be connected with the presence of 
3d impurities in the $Ca$ precursor. 
\par
For the $1000~cm^{-1}$ group, it is sometimes described by four components, 
or even more. We consider that, as only three bumps are unambiguously 
visible on the spectra, the lowest number of components giving 
satisfactory reconstruction of the experimental data is the best 
possible choice. The four components needed by Kalampounis et al. \cite{r11}
look obviously necessary on their data for $x>0.55$ and $x<0.40$, but not in
the range investigated here. 
\par
Modes 2-3 are somewhat weak and large, and then are rather inaccurate 
(at least compared to other ones) as their possible parameter variation 
in the $Ca$ range is hindered by experimental uncertainty. Mode 1 is 
discussed below. For the 
following modes $4$ and $5$, the accuracy is better. Both wavenumbers
monotonously increase upon increasing $Ca$ concentration (see Figure 
\ref{raman2}). Their intensities 
show a transfer from $5$ to $4$ somewhat above $47\%$ $Ca$. 
For linewidths, mode $4$ 
smoothly broadens whereas mode $5$ suddenly narrows above $50\%$. 
The greater effects concern the higher frequency modes, $6$, $7$, $8$ and $9$. 
One can first consider the latter three modes, clearly connected in 
one broad band. This assignement slightly differs from \cite{r11}, where mode
7 is assigned to $Q^0$ species, and from Frantz and Mysen \cite{Mysen} where
$Q^3$ corresponds to a weaker mode and mode 8 is assigned to Si-bonding 
oxygen stretching vibrations. The superscript 
$n$ in $Q^n$ denotes the number of bridging oxygens (not connected to Ca) on a 
$SiO_{4/2}$
tetrahedron.According to Zotov \cite{Zotov}, mode $9$ 
can be assigned  
to $Q^3$ species, mode $8$ to $Q^2$ and mode $7$ to $Q^1$. 
The wavenumbers for 
modes $7$ and $8$ are practically constant. It is not the case of their 
intensities, which show a continuous increase for modes $7$ and $8$, 
whereas mode $9$ is rather constant up to $47\%$ $Ca$, and decreases rapidly 
after. The wavenumber of mode $9$ displays a decrease up to around 
$47\%$ $Ca$, 
and then a slight increase. This effect is small but clearly larger 
than experimental error. One can note that the sum of intensities of 
modes $8$ and $9$ is roughly constant above this value $47\%$ $Ca$
(Figure \ref{raman1}). For 
linewidths, modes $7$ and $8$ are slightly increasing, whereas mode $9$ 
is constant up to $47\%$ $Ca$, and shortens after (slightly, but 
clearly above experimental error). 
One can then conclude on this part that the increase of $Ca$ acts in 
two ways : a first one, continuous, leading to increase the number of 
$Q^2$ and $Q^1$ species, to equilibrate the higher number of cations, and 
a second one, sharper, which consists in a transfer 
from $Q^3$ species 
to $Q^2$ ones upon increasing $Ca$ content above $47\%$. Then the remaining 
$Q^3$ are more decoupled of the network, as shown by their lower linewidth. 
\par
Mode $6$ is the most affected by the modification of $Ca$ content as shown
on Fig. \ref{raman1}. Its 
frequency is hugely increased (more than $50~cm^{-1}$, which is considerable), 
the width of the line is decreased by more than a factor $2$, between
 $42$ to $48\%$ calcium, and its 
integrated intensity falls down by one order of magnitude in the same 
concentration range. When comparing with Ref. \cite{r11} where the 
$600~cm^{-1}$ asymmetric component is described only by one line, the present 
work shows that the anomaly standing around $x=47\%$ is only due to mode 6, the
other modes 4 and 5 overlapping in this component evolve monotonously in this
composition range. This mode 6 cannot then be assigned to $Q^0$ 
species as the corresponding intensity must increase upon enriching $Ca$ 
content. It is more plausibly connected with $Si-O-Si$ motions.  
The intensity dependence which is more rapid than 
for mode 9, would lead to think that mode $6$ could correspond to 
$Si-O-Si$ vibrations of $Q^3-Q^3$ species. 
\begin{figure}
\begin{center}
\includegraphics[width=0.9\linewidth]{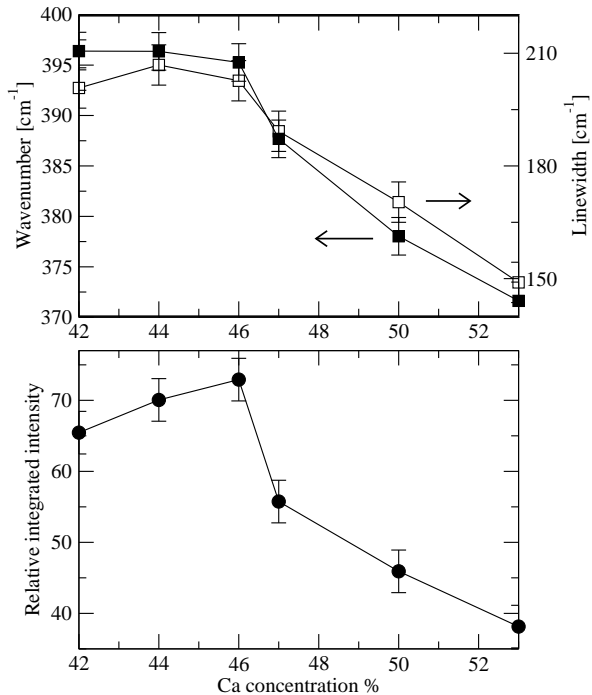}
\caption{\label{raman3} Compositional behaviour in relative wavenumber, 
linewidth and relative integrated intensity of the mode 1 line lying at 
$392~cm^{-1}$ for $42\%$ Ca.}
\end{center}
\end{figure}
\par 
Mode 1 displays two specific features around $46-47\%$  Ca. Its frequency 
exhibits a noticeable downshift ($20~cm^{-1}$), the width decreases by $30\%$, 
and the integrated intensity falls down by a factor two (Figure \ref{raman3}).
Even if this mode 
is broader than the preceding ones, these parameter evolutions are 
sufficiently large to be unambiguous.  This frequency downshift and line 
narrowing upon increasing Ca content is different from high-frequency modes 
(6, 9). This softening and narrowing is an uncommon behavior and can 
be qualitatively explained only by a significant change of eigenvectors. 
Due to its frequency (it is the lowest-frequency mode apart from the 
boson peak), it implies some Ca vibrations. These vibrations would then 
more decoupled, and softer, above ca. $46\%$, that can be attributed to 
floppy regions in the glass.
\subsection{Raman threshold and rigid to floppy transitions}
The results obtained from Raman scattering and the behaviour with 
concentration of some modes closely parallels the 
one found in chalcogenides \cite{PRB2000}. The frequency increase exhibited 
by modes $6$ and $9$ above $47\%$ calcium can indeed be discussed in 
the context of rigid to floppy transitions. In fact, these modes look to 
harden just when the network becomes softer. This apparent  contradiction 
can be explained by a non-uniform distribution of stress : in the stressed 
rigid range close to our observed threshold at $x=0.47$, 
the network is mainly hardened by interconnected $Q^3$ species. 
Upon undergoing a rigidity transition, these links should become much less 
numerous, leading to a network which is dominated by $Q^2$ connections. The 
remaining $Q^3$ islands should become somewhat decoupled from this network, 
producing an increase of the mode 9 wavenumber and a noticeable line 
narrowing : the distribution of $Q^3$ force constants is less important, 
leading to a 
linewidth closer to a lorentzian one. In the floppy range at high 
calcium concentration, the structure should be a 
network of $Q^2$ elements, separated by small 'harder' islands. The narrowing 
exhibited by mode $5$ above $50\%$  can be also interpreted in the same way.
\section{Analysis from constraint theory}
Maxwell constraints counting appears to be useful to understand the present
results as the calcium silicate network can be described by a molecular
system constrained by bond-stretching and bond-bending (angular) forces.
\subsection{Maxwell global constraint counting}
We consider the $CaO-SiO_2$ system as a network of $N$ atoms
composed of $n_r$ atoms that are $r-$ fold coordinated. Enumeration of
mechanical constraints \cite{r15}-\cite{r17}
associated with bond-stretching (radial) forces
leads to $n_c^\alpha=r/2$, while the number of bond-bending (angular) 
constraints is $n_c^\beta=(2r-3)$. The average number of floppy modes per atom 
$F/N$ in this
three-dimensional network is given by\cite{r17}:
\begin{eqnarray}
\label{e1}
f=F/N&=&n_d-n_c=3-{\frac {1}{N}}\sum_{r\geq 2}n_r({\frac {5r}{2}}-3)
\end{eqnarray}
where $n_d$ is the dimension of the network and $n_c=n_c^\alpha+n_c^\beta$. 
Applied to the system of 
interest leads to:
\begin{eqnarray}
\label{F/N}
f=F/N&=&3-{\frac {11-7x}{3-x}}
\end{eqnarray}
The latter equation holds if one assumes that silicon is four-fold,
calcium and oxygen are two-fold
coordinated \cite{RKJCP}. 
As one can see, the number of floppy modes vanishes when the
network attains the critical concentration $x=x_c=0.50$ which is in close
agreement with the thresholds observed from our Raman results. 
\par
Equation (\ref{F/N}) defines a mean-field transition in which the number 
of mechanical constraints is computed from the macroscopic concentration $x$. 
Obviously,
this elastic transition may be attained at $x<x_c$ provided that some 
macroscopic floppy subregions can emerge with the addition of alkaline 
earth oxide. On the other hand, it has been recently shown 
\cite{r20},\cite{PRB2003} that the 
underlying nature of the floppy to rigid transition was more subtle and could
contain under certain circumstances \cite{r19} two 
transitions instead of the single 
one predicted by mean-fied constraint counting. In this context, stress will 
not spread randomly over the whole network as initially believed but will
accumulate in underconstrained subregions leading to the occurence of an
intermediate elastic phase that is found to be stress free. The 
detection of this intermediate phase has been mostly accomplished from
calorimetric probes using temperature modulated differential scanning 
calorimetry (MDSC) \cite{r27}. Here, a sinusoidal variation is added on the
usual linear DSC ramp and permits to decompose the total heat flow into a 
reversing part that tracks the initial modulation and a residue. The latter is
found to vanish in the intermediate phase \cite{r19}.
\subsection{Cluster construction}
Size increasing cluster approximations (SICA) appear to be a useful tool to
describe the elastic nature of the network backbone (floppy, intermediate,
stressed rigid) and permits to take into account medium-range order effects
that may, or may not, serve as ingredient for the presence of an
intermediate phase. Furthermore, SICA start from the mean-field description as
basic level and takes into account non-random structural elements and
their related mechanical constraints in a systematic fashion.
This method has been first introduced to study the formation of
fullerenes \cite{C60} or Penrose tilings in quasicrystals \cite{quasi} and
was then applied to quantify the boroxol ring
statistics in amorphous $B_2O_3$ \cite{Dina}. Recently, SICA has been used in
the context of floppy to rigid transitions \cite{PRB2003} in an archetypal
chalcogenide network ($Ge_xSe_{1-x}$) and has led to the
definition of a stress-free intermediate phase that depends substantially
on the fraction of small rings in the structure.
\par
The basic level ($l=1$) of the SICA construction corresponds to the mean-field
approximation, having as elements structural species that depend
directly on the macroscopic concentration. The construction permits to
generate clusters at step $l=2$ sharing all possible combination of the
basic ($l=1$) elements, clusters at step $l=3$ having three ($l=1$) units,
etc. The probabilities of the clusters are computed within the 
Canonical Ensemble having energy levels $E_n$ related to bond creation
between the basic level molecules. The construction is furthermore supposed
to be performed at the
formation of the network when T equals the fictive temperature $T_f$ 
which is defined by the intersection of the extrapolated supercooled liquid and glass curves \cite{Galeener}.
Mathematically, these probabilities will involve statistical weights (or
degeneracies $g(E_n)$ of a corresponding energy level) which correspond to
the number of equivalent ways to connect two ($l=1$) basic units, and
a Boltzmann factors \cite{Galeener} of the form $e_n=exp[-E_n/T_f]$.
\par
We have used as basic units the $CaO$ molecule and a $SiO_{4/2}$
tetrahedron, having for respective probabilities the macroscopic
concentration $x$ and $(1-x)$. 
The energy levels are defined from the consideration of all
possible connections at the step $l=2$ and permit to distinguish the
mechanical nature of the underlying clusters (floppy, isostatic,
stressed).  The creation of
a chain-like $Ca_2O_2$ floppy structure ($n_c<3$) is related to an
energy gain of $E_f$ while the creation of a CaO connected to a
silicon tetrahadron (an isostatically rigid $CaSiO_3$ cluster, $n_c=3$) is
associated with
an energy $E_i$. Finally, the basic network former represented by two
connected $SiO_{4/2}$ tetrahedra (a stressed rigid $Si_2O_4$ cluster,
$n_c>3$) corresponds to an energy gain of $E_s$. 
At step $l=2$, three different clusters can be
obtained (see fig. \ref{construct}) 
and their probabilities are given below. 
\begin{figure}
\begin{center}
\includegraphics[width=0.5\linewidth]{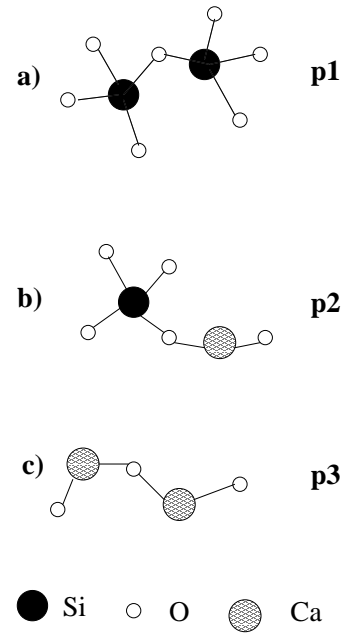}
\caption{\label{construct} The different clusters obtained at SICA step
$l=2$. a) A fragment of the silica network with $g(E_s)=16$, b) a 
wollastonite-like cluster $CaSiO_3$ with $g(E_i)=16$ and c) a calcium-rich 
cluster (c) with $g(E_f)=4$.}
\end{center}
\end{figure}
We
mention that edge-sharing $SiO_{4/2}$ have been excluded from the
construction as they have no experimental evidence at all in silicates
\cite{r29}. The probabilities at setp $l=2$ are:
\begin{eqnarray}
\label{p1}
p_1&=&{\frac {16(1-x)^2e_s}{4x^2e_f+16x(1-x)e_i+16(1-x)^2e_s}}
\end{eqnarray}
\begin{eqnarray}
\label{p2}
p_2&=&{\frac {16x(1-x)e_i}{4x^2e_f+16x(1-x)e_i+16(1-x)^2e_s)}}
\end{eqnarray}
\begin{eqnarray}
\label{p3}
p_3&=&{\frac {4x^2e_f}{4x2e_f+16x(1-x)e_i+16(1-x)^2e_s}}
\end{eqnarray}
\par
out of which can be computed the ($l=2$) concentration of calcium atoms 
\begin{eqnarray}
\label{conc}
x^{(2)}&=&{\frac {p_2+2p_1}{4-p_2-2p_1}}
\end{eqnarray}
Due to the initial choice of the basic units, the energy $E_i$ will
mostly determine the probability of isostatic clusters since the
related Boltzmann factor $e_i$ is involved in the probability
(\ref{p2}) of creating the isostatic $CaSiO_3$ cluster (a $CaO-SiO_{4/2}$
bonding). In the case where $E_i\ll E_f,E_s$, the bonding of the network 
construction will be mainly achieved by isostatic clusters. 
\begin{figure}
\label{cluster}
\begin{center}
\epsfig{figure=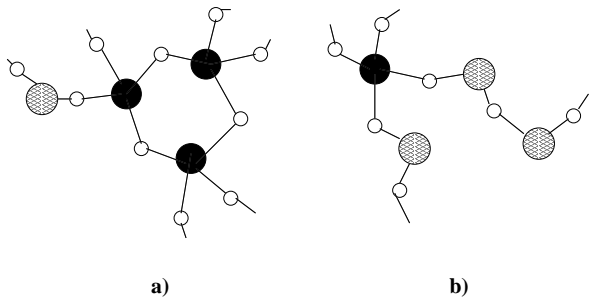,width=0.9\linewidth}
\end{center}
\caption{\label{cluster} Clusters generated at SICA step $l=4$. a) 
a 3-membered ring
at low Ca concentration. b) A high Ca structure.}
\end{figure}
\par 
For larger steps ($l>2$), one has to take care of possible isomers produced 
by two distinct clusters at the lower level. The cluster displayed 
in Fig \ref{cluster}a can for example be produced either by connecting
two $SiO_{4/2}$ tetrahedra onto a $CaSiO_3$ ($l=2$) cluster or
by adding $CaO$ and $SiO_{4/2}$ onto a $Si_2O_4$ ($l=2$) cluster. 
The statistical weight $g(E_n)$ of such 
isomers will be of course larger than the one of low connected clusters having a single pathway of production.
\par
As seen from equs. (\ref{p1})-(\ref{p3}), the cluster probabilities will 
depend only on two energetical parameters (i.e. the factors 
$e_s/e_i$ and $e_f/e_i$) and the problem can still be reduced by using a charge
conservation law \cite{Bray} for the calcium atom 
applied on the population of clusters of size $l$.
\begin{equation}
\label{1}
x^{(l)}=x
\end{equation}
The $x$-dependence of either $e_s/e_i$ or
$e_f/e_i$ means that either the fictive temperature $T_f$ or the energies
$E_n$ depend \cite{Galeener} on $x$ but here only the $e_n(x)/e_i(x)$
dependence is relevant for the analysis.  
\par
In order to obtain some clusters having significant medium range order, 
the SICA construction has been realized up to the step $l=4$.
\subsection{Maxwell cluster constraint counting}
Next, one can apply on the generated set of clusters Maxwell constraint
counting by enumeration of bond-bending and bond-stretching 
constraints and calculation of the 
corresponding expressions of $n_c^\alpha$ and  $n_c^\beta$. Of particular
importance are the structures containing a ring having less than six atoms
(see Fig. \ref{cluster}a), because one has to remove some extra constraints
\cite{r16}. 
\par
For each step $l$ has been computed the total number of constraints $n_c^{(l)}$:
\begin{eqnarray}
\label{ncl}
n_c^{(l)}={\frac {\sum_{i=1}^{{\cal N}_l}n_{c(i)}p_i}{\sum_{i=1}^{{\cal N}_l}N_ip_i}}
\end{eqnarray}
where ${{\cal N}_l}$ is the total number of clusters of
size $l$ and $n_{c(i)}$ and $N_i$ are respectively the number of
constraints and the number of atoms of the cluster of size $l$ with 
probability $p_i$. 
Applied to the set of clusters at step $l=2$, one obtains for example:
\begin{eqnarray}
\label{2b}
n_c^{(2)}&=&{\frac {22p_1+15p_2+4p_3}{6p_1+5p_2+2p_3}}
\end{eqnarray}
We have determined either
$e_s/e_i$ or $e_f/e_i$ as a function of concentration, by solving the charge 
conservation law (\ref{1}). With these factors depending on the concentration 
$x$, it is possible to compute the cluster probabilities $p_i$ of a given step $l$ as a function of composition and finally obtain the composition $x_c$ 
where the number of floppy modes $f_l=3-n_c^{(l)}$ vanishes. It can be also 
computed the statistics of $Q^n$ species with Ca concentration as 
discussed next.
\section{Results}
\subsection{Structural properties and speciation}
In this section, we consider the solutions of the SICA construction in terms of structure. One principal objective of the
present investigation is first to compare the model results 
with some experimental data on calcium silicate glasses 
such as the relative abundances
and mixing properties of the structural units with respect to the 
concentration.
\par
\begin{figure}
\begin{center}
\includegraphics[width=0.9\linewidth]{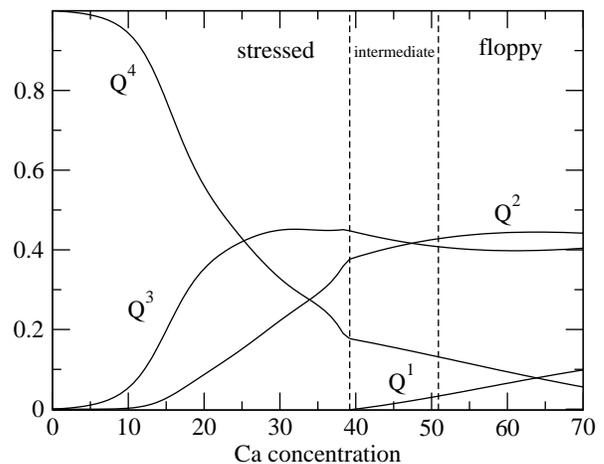}
\end{center}
\caption{\label{Qi} $Q^n$ distribution in $(1-x)SiO_2-xCaO$ as a 
function of the calcium 
concentration, computed at SICA step $l=4$. The vertical broken lines
show the boundaries of the intermediate phase (see below).}
\end{figure}
\begin{figure}
\begin{center}
\epsfig{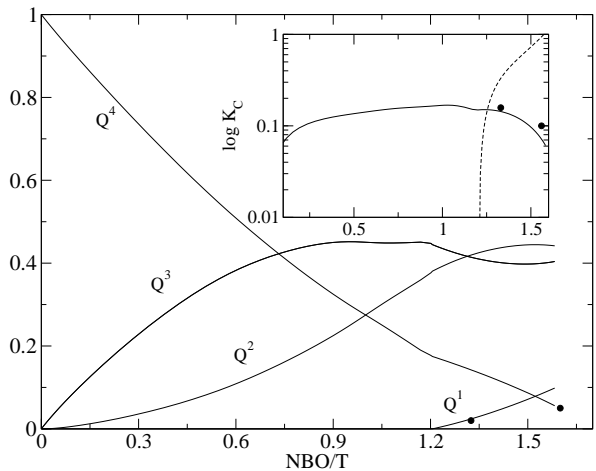}
\end{center}
\caption{\label{nboT}$Q^n$ distribution 
in $(1-x)SiO_2-xCaO$ as a function of NBO/T, 
computed at SICA step $l=4$. The insert shows the calculated
equilibrium constant 
$K_{C1}$ (solid line) and $K_{C2}$ (broken line) as a function of NBO/T. 
See text for details. Data points are taken from Frantz and Mysen [29].}
\end{figure}

Figures \ref{Qi} and \ref{nboT} represent the distribution of $Q^n$ units 
computed from SICA at step $l=4$, respectively as a function of Ca 
concentration and as a function of the relative abundance of 
non-bridging oxygens NBO/T
that is defined by:
\begin{eqnarray}
\label{nbo}
\sum_kn_kx(Q^k)=NBO/T
\end{eqnarray}
where $n_k$ is the fraction of non-bridging oxygens per $SiO_{4/2}$ tetrahedra
 and  $x(Q^k)$ is the concentration of $Q^k$ species. The quantity defined by 
equation (\ref{nbo}) is commonly used to quantify network depolymerization 
with non-bridging oxygens \cite{r30}.
\par
As one can see, the probability of finding a $Q^4$ unit decreases smoothly 
with the addition of calcium whereas the emergence of $Si_2O_5^{2-}$ ($Q^2$)
and $SiO_3^{2-}$ ($Q^3$) groupings is noticeable. One can furthermore remark the
crossover of the abundance of $Q^2$ and $Q^3$ at the concentration $x=0.47$, 
a behaviour that has been also observed from the integrated intensity of mode 
8 and 9 in the Raman spectra. 
\par
Several limitations of the approach can however be objected. First, the fact 
that we limit the cluster construction to the step $l=4$ does not permit to
generate $Q^0$ as these species are made of a central tetrahedron that shares
four Ca atoms (that structure would be created by $l=5$). As a consequence, 
the  distribution of our $Q^n$ ($n=4,3$) are slightly overestimated in the high 
calcium range. 
Secondly, as the energetical factors are introduced only on the basis of the
overall mechanical character of the bonding type (stressed, isostatic, floppy)
between two basic units, difficulties arise in order to describe the 
behaviour with changing alkaline earth 
cation as the size (or energetics, or steric hindrance,...)
does not appear in this approach. This means that other ingredients 
\cite{RKJCP} are 
necessary to describe the differences arising in Magnesium or Barium silicates.
\par
Next, we can focus on equilibrium constant \cite{r31} $K_{c1}$ and $K_{c2}$
between species respectively related to the equilibria:
\begin{eqnarray}
\label{equi1}
Si_2O_5^{2-}\rightleftharpoons SiO_2+SiO_3^{2-} 
\end{eqnarray}
\begin{eqnarray}
\label{equi2}
4SiO_3^{2-}\rightleftharpoons Si_2O_5^{2-}+Si_2O_7^{6-}
\end{eqnarray}
The mass action constants can be computed from the above equilibria using the
computed distribution of $Q^n$ species. It is obvious that equ.(\ref{equi1}) 
will be the dominant equation in the low calcium 
region while both equilibria will
have to be taken into account in the concentration range lying around the 
rigid to floppy transition when NBO/T is larger than the value $1$. 
As one can see 
from the insert of Fig. \ref{nboT}, the calculated $K_{c1}$ from the 
equilibrium constant of (\ref{equi1}) is in fair agreement with the two data 
points number reported by Mysen \cite{Mysen}. We note also that $K_{c1}$ remains
almost constant over the entire concentration range of interest suggesting that 
the conversion between species is not favoured in the stressed rigid side of 
the glass formation range. On the other hand, the rapid variation of $K_{c2}$ 
can arise from the floppy nature of the backbone, when $NBO/T>1.25$. The change
in equilibrium constant following the underlying elastic nature of the
network has been questioned by Eckert and co-workers in phosphorus 
chalcogenides \cite{Eckert}.
\subsection{Rigidity transitions from cluster probabilities}
From the constraint counting applied on the set of cluster of a given size 
$l$, it is possible to extract a certain number of quantities of interest in
the context of rigid to floppy transitions. 
\par
\begin{figure}
\begin{center}
\epsfig{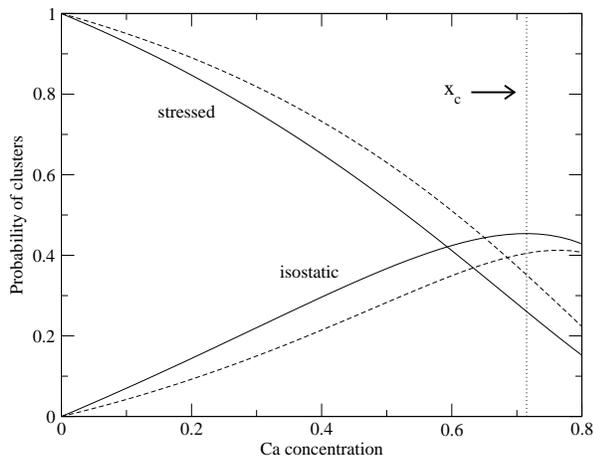}
\end{center}
\caption{\label{probclustrando} Probability of finding isostatically rigid and
stressed rigid clusters as a function of calcium concentration $x$ 
for SICA steps $l=3$ (soldi line) and $l=4$ (broken line)
in the case of random bonding. The vertical line corresponds to the 
mean-field rigidity transition $l=4$ defined by $f=0$.}
\end{figure}

The simplest case allowed in the case of the SICA construction is the case
of random bonding when no selective rules are given on the energies to retain
specific clusters as starting point for the next cluster generation. 
This means hat the cluster probabilities 
$p_n$ are only given by their statistical weights $g(E_n)$ and reduces for 
instance the probability of a $CaSiO_3$ cluster (see equ. (\ref{p2}))
at step $l=2$ to:   
\begin{eqnarray}
\label{p4r}
p_2=p_{CaSiO_3}&=&{\frac {x(1-x)}{(2-x)^2}}
\end{eqnarray}
or the probability $p_{3R}$ of a three-membered ring $Si_3O_6$ at $l=3$ to:
\begin{eqnarray}
\label{p3r}
p_{3R}={\frac {36(1-x)^3}{(11-10x)(2-x)^2}}
\end{eqnarray}
It is found (see Table \ref{table2}) that the vanishing of the number 
of floppy modes $f$ occurs at a somewhat higher value $x_c$ than 
the macroscopic Maxwell value of $0.50$. Also, if one considers only dendritic
clusters (i.e. no allowance for ring structures), the threshold $x_c$ is 
lowered for steps $l=3$ and $l=4$. No change occurs for step $l=2$ as there are
no rings created at this step. The lowering of $x_c$ for dendritic clusters 
arises from the fact that the stressed rigid ring structures which have a high
statistical weight $g(E_n)$ due to their higher connectivity, are now absent
leading to a number of constraints computed from equ. (\ref{ncl})
that is lowered. 
\begin{table}[t]
\caption{\label{table2} Glass optimum condition $f=0$ for different SICA steps and different cluster types in the case of random bonding.}
\begin{ruledtabular}
\begin{tabular}{lcccc}
SICA step $l$&1&2&3&4 \\
$x_c$ &0.500&0.739&0.690&0.715 \\
$x_c$ (dendritic)&0.500&0.739&0.667&0.681 \\
\end{tabular}
\end{ruledtabular}
\end{table}
The probability of finding floppy, isostatically rigid and stressed rigid 
clusters can be computed for each SICA step and it is found that for 
dendritic clusters there is a global increase of the floppy and 
isostatically rigid clusters with increasing Ca concentration whereas the 
probability of stressed rigid clusters is continuously decreasing
(see Figure \ref{probclustrando}).
However, no threshold is observed as the one suggested by the 
Raman analysis and one obtains a single solution $f=0$ corresponding 
to the glass optimum point defined by the vanishing of the number 
of floppy modes.
\subsection{Boolchand intermediate phase}
We now turn to self-organization that permits to obtain an elastic 
intermediate phase \cite{r20},\cite{PRB2003}. This elastic phase has 
been first observed by Boolchand and co-workers in chalcogenides \cite{r19},
but as rigid to floppy transitions have been also detected in oxide glasses
 \cite{Science}, there is no reason why this 
intermediate phase should not exist in the present system as well. We consider
here the strutural possibilities that can lead to an intermediate phase.
It is achieved with SICA 
by selecting the pathways of cluster production. One starts for example
with a underconstrained (floppy) cluster of size $l$ 
which exists at high calcium
concentration. Agglomeration of basic units $l=1$ onto this cluster is
only allowed if the creation of a stressed rigid region can be avoided 
on the generated cluster of size $(l=1)$. This would happen if one starts
to connect two $SiO_{4/2}$ tetrahedra together, involving an energy gain of 
$E_s$. On the other hand, if $x$ is decreased one will accumulate with 
this rule isostatically rigid regions on the generated clusters as the 
only allowed connections are either floppy $CaO-CaO$ or 
$CaO-SiO_{4/2}$ bondings. Alternatively, if one starts from the low
calcium side, self-organization can be obtained by selecting along the
same scheme stressed rigid and isostatically rigid connections and 
excluding systematically the possibility of floppy $CaO-CaO$ bondings in
the SICA construction. 
\par
With decreasing calcium content, one will be able to maintain that rule up
to a certain point beyond which connections between two silicon tetrahedra
(or a stressed rigid connection) can not be avoided anymore. The latter point
corresponds to a stress transition \cite{PRB2003} and appears only if
some medium range order (MRO) made of rings is accepted in the construction.
The concentration range bounded on its low calcium side by 
the stress transition and on its high calcium side by the vanishing of
the number of floppy modes, defines the intermediate phase of width $\Delta x$.
\begin{figure}
\begin{center}
\includegraphics[width=0.9\linewidth]{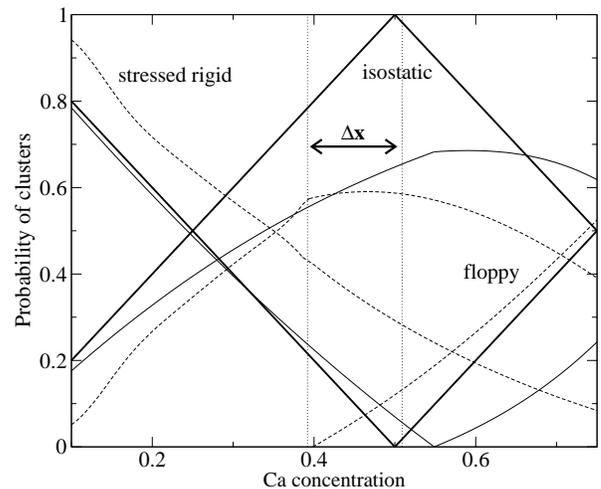}
\end{center}
\caption{\label{probcluster} Probability of finding stressed rigid, isostatically rigid and
floppy clusters as a function of calcium concentration $x$. The thick
solid line corresponds to the basic SICA step $l=1$ and $l=2$, 
the thin solid line to $l=3$ and the broken line to $l=4$. The vertical lines
define the intermediate phase of width $\Delta x$ at setp $l=4$. The present situation 
for $l=3$ and $l=4$ corresponds to self-organization with ring structures.
Notice the kink around $x=0.40$ corresponding to the stress transition (see 
text for details). }
\end{figure}
Results of self-organization are displayed in Fig. \ref{probcluster}.
The simplest case for self-organized clusters is again the case where 
rings are removed from the construction corresponding to dendritic clusters
which would correspond in the limit $l\to \infty$ to Bethe lattice solutions 
\cite{RBM} or Random Bond Models \cite{Bethe} obtained in the context of
floppy to rigid transitions.
We obtain here a single transition for all SICA steps $l$ either at the
mean-field value $x=x_c=0.50$ or closeby ($x_c=0.509(7)$ for $l=4$). No 
intermediate phase is obtained.
The probabilities of floppy, isostatically rigid and stressed
rigid clusters can be displayed as a function of calcium content
(Fig. \ref{probcluster}) and show that the abundance of isostatically
rigid clusters is maximum at the threshold defined by $f=0$. This is 
obviously the case for the $l=2$ case, but also for larger SICA steps.
\par
\begin{figure}
\begin{center}
\includegraphics[width=0.9\linewidth]{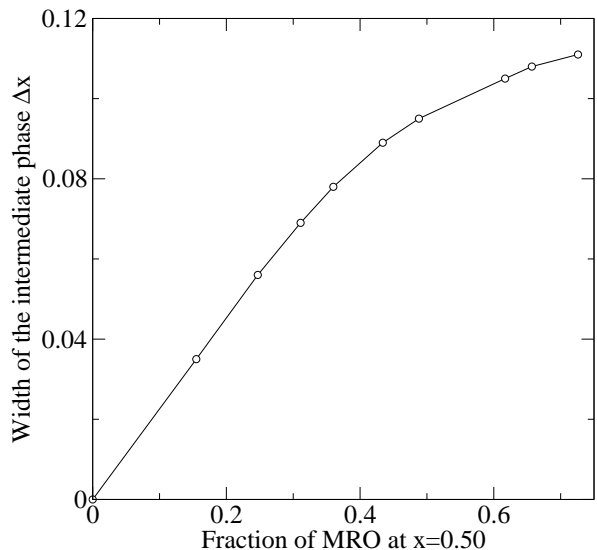}
\end{center}
\caption{\label{width} Width of the intermediate phase at step $l=4$ 
in $(1-x)SiO_2-xCaO$ glasses, as a function  of the fraction 
of medium range order (rings) computed at $x=0.50$. The insert represents the
computed number of floppy modes for step $l=3$ (solid line) and $l=4$ 
(broken line).} 
\end{figure}
The intermediate phase shows up if a certain amount of medium range
order (MRO) is allowed. This is realized in the SICA construction by
generating in the construction ring clusters such as the one displayed in
Fig. \ref{cluster}. The requirement of self-organization in the cluster
construction still holds for dendritic stressed rigid structures made of 
at least two connected $SiO{4/2}$ tetrehadra which propagate stress in 
the structure. But now cyclic structures such as rings are preserved from
self-organization.
Two transitions are obtained for every SICA steps $l>2$ (see Fig. 
\ref{probcluster}. 
The first transition one lies always around the concentration $x_{c2}=0.50$ 
calcium
and corresponds to a rigidity transition where the number of floppy modes
vansihes. The second transition that emerges with increasing MRO is located
for $l=4$ at $x_{c1}=0.392$ and corresponds to the stress transition. When
starting from a floppy network at high calcium concentration, the progressive
stiffening of the network can be accomplished by requirement of
self-organization leading to the accumulation of 
isostatically rigid regions and stressed rigid ring structures. This will work
for any decreasing $x$ up to $x_{c2}$ below which stressed 
rigid bondings outside of ring structures can not be avoided anymore. 
\par
The two transitions $x_{c1}$ and $x_{c2}$ will define an intermediate phase 
$\Delta x$ that depends on the fraction of MRO allowed and we show that $\Delta x$ 
is anincreasing function of the MRO (Fig \ref{width}). Furthermore, as
the rigidity transition $f=0$ at $x_{c2}$ is almost not affected by 
the presence of MRO, the increase of the width with the latter quantity arises
mostly from the decrease of $x_{c1}$ with growing MRO.
Finally, as seen from Fig. \ref{probcluster}, there is kink observable at 
$x_{c1}$ which would produce in first-order derivatives such as the energy
\cite{Naumis} a jump suggesting a first-order stress transition at $x_{c1}$
\cite{r20},\cite{PRB2003} and a continuous rigidity transition at$x_{c2}$.
\par
The elastic nature of the network can be also analyzed within this framework. 
Fom Fig. \ref{probcluster}, one can see that the probability of finding
isostatically rigid clusters is maximum in the window $\Delta x$. It is equal 
to 1 for the $l=2$ SICA step and about $0.5$ or $0.6$ for the larger steps thus
providing evidence that the molecular structure of the network in the 
window is almost stress-free. 
\section{Summary and conclusions}
We have shown in the present work from Raman measurements that a particular 
transition was observed at around $47\%$ calcium in the calcium silicate 
system, in harmony with previous findings \cite{r11}. We have elucidated 
the origin
of the observed threshold as being the manifestation of a global softening
of the glass structure that signals ultimately in a rigid to 
floppy transition. Constraint counting algorithms applied on size 
increasing clusters permit to refine the picture by predicting 
the possibility of an intermediate phase in the Raman threshold region.
\par  
Recent calorimetric and spectroscopic studies have shown that 
this kind of elastic transitions 
could be found in alkaline silicates as well, with a well-defined signature
of the intermediate phase in both sodium \cite{Vaills2004} and potassium 
glasses \cite{unpub}. We are confident that a similar generic behaviour should
be expected in the alkaline earth silicate glasses.
A definite probe of the existence of the intermediate phase in the present
system could be obtained using MDSC measurements during the glass transition.
The results present always a
vanishing of the kinetic dependent heat flow inside that phase \cite{r19} and
provide an unambigous signature for both transitions, rigidity and stress. 
Unfortunately, the glass transition temperatures of calcium silicates are too high 
to be accepted from the actual performances of the MDSC apparatus\cite{MDSC}.

\section*{ACKNOWLEDGMENTS}
Discussion and comments with P. Boolchand are gratefully acknowledged. 
We thank Emmanule Veron for the EDX analysis and Boris Robert for 
help during the cours of this work. LPTL is   Unit\'e Mixte de Recherche  du
CNRS n. 7600. CRMHT is Unit\'e Propre du CNRS n. 4212.

\end{document}